\documentclass[10pt, letterpaper]{article}
\usepackage{amsfonts,amsmath,amssymb,mathrsfs,verbatim}

\let\n\noindent

\newcommand{\btab}{\begin{tabular}}     \newcommand{\etab}{\end{tabular}}

\newcommand{\bt}{\begin{table}}     \newcommand{\et}{\end{table}}

\newcommand{\ba}{\begin{array}}     \newcommand{\ea}{\end{array}}
\newcommand{\bc}{\begin{center}}        \newcommand{\ec}{\end{center}}
\newcommand{\bfig}{\begin{figure}}      \newcommand{\efig}{\end{figure}}
\newcommand{\bp}{\begin{picture}}       \newcommand{\ep}{\end{picture}}
\newcommand{\bq}{\begin{quote}}     \newcommand{\eq}{\end{quote}}
\newcommand{\ben}{\begin{enumerate}}    \newcommand{\een}{\end{enumerate}}

\font\tenmsy=msbm10
\font\sevenmsy=msbm10 at 7pt
\font\fivemsy=msbm10 at 5pt
\newfam\msyfam 
\textfont\msyfam=\tenmsy
\scriptfont\msyfam=\sevenmsy
\scriptscriptfont\msyfam=\fivemsy
\def\blackB{\fam\msyfam\tenmsy}
\def\Z{{\blackB Z}}

\let\R\rangle

\def\frac#1#2{{\textstyle{#1\over #2}}}

\def\text#1{\quad\hbox{#1}\quad}
\def\gh{\hat{g}}

\def\la{\lambda}
\def\e{\epsilon}

\def\muh{{\hat \mu}}

\def\A{{\cal{A}}}

\def\lah{{\hat \lambda}}

\def\om{{\omega}}
\def\y{{\infty}}

\def\gh{{\widehat g}}
\def\rw{\rightarrow}

\def\R{\rangle}

\def\su{\widehat{su}}
\def\osp{\widehat{osp}}

\def\frac#1#2{{#1 \over #2}}

\def\suh{{\widehat {su}}}

\def\M{{\cal {M}}}
\def\uh{{\widehat u}}
\def\rw{{\rightarrow}}

\let\R\rangle

\def\frac#1#2{{\textstyle{#1\over #2}}}

\def\text#1{\quad\hbox{#1}\quad}
\def\gh{\hat{g}}

\def\la{\lambda}
\def\e{\epsilon}

\def\muh{{\hat \mu}}

\def\A{{\cal{A}}}

\def\lah{{\hat \lambda}}

\def\om{{\omega}}
\def\y{{\infty}}

\def\gh{{\widehat g}}
\def\rw{\rightarrow}

\def\R{\rangle}

\def\su{\widehat{su}}
\def\osp{\widehat{osp}}

\def\frac#1#2{{#1 \over #2}}

\def\suh{{\widehat {su}}}
\def\uh{{\widehat u}}
\def\rw{{\rightarrow}}

\begin{document}

\vskip18pt

\title{\vskip60pt The $Z_k^{(su(2),3/2)}$ parafermions}

\vskip18pt

\smallskip
\author{ P. Jacob and P.
Mathieu\thanks{patrick.jacob@durham.ac.uk,
pmathieu@phy.ulaval.ca.} \\
\\
Department of Mathematical Sciences, University of Durham, Durham, DH1 3LE, UK\\
and\\
D\'epartement de physique, de g\'enie physique et d'optique,\\
Universit\'e Laval,
Qu\'ebec, Canada, G1K 7P4.
}

\vskip .2in
\bigskip
\date{June 2005}

\maketitle


\vskip0.3cm
\centerline{{\bf ABSTRACT}}
\vskip18pt

We introduce a novel parafermionic theory for which the conformal dimension of the basic parafermion is $\tfrac32(1-1/k)$, with $k$ even.
The structure constants and the central charges are obtained from
mode-type associativity calculations. The spectrum of the
completely reducible representations is also determined. The
primary fields turns out to be labeled by two positive integers
instead of a single one for the  usual parafermionic models.  The
simplest singular vectors are also displayed. It is argued that
these models are equivalent to  the non-unitary minimal
$W_k(k+1,k+3)$ models. More generally, we expect all
$W_k(k+1,k+2\beta)$ models to be identified with generalized
parafermionic models whose lowest dimensional parafermion has
dimension $\beta(1-1/k)$.



\n



\newpage


\section{Introduction}

The Fateev-Zamolodchikov $Z_k$ parafermionic conformal field
theory, originally  introduced in \cite{ZFa}, can be generalized
along two quite distinct  lines. The first one relies on the
observation that the standard parafermionic model is equivalent to
the coset  $\suh(2)_k/\uh(1)$. In that vein, a  natural
generalization amounts to consider the cosets $\gh/\uh(1)^r$ where
$r$ is the rank of the affine Lie algebra $\gh$ \cite{Gep}.
Originally formulated in terms of untwisted affine simple  Lie
algebras, this approach has  been further extended  to the cases
where $\gh$ is either twisted \cite{DGZ} or contains fermionic
generators \cite{CRS}.

For the second line of generalization, one   preserves the  structure of the standard
 theory, i.e., the cyclic structure  of the OPEs (namely $\psi_n\times \psi_m\sim \psi_{n+m}$ with  $\psi_k\sim I$),  but attributes  different conformal dimension to the parafermionic fields. We recall in that regard that for a candidate chiral algebra whose OPE has a $Z_
k$ invariance,  the dimension of the basic parafermionic fields,
$h_{\psi_n}$, have to be introduced as an input.  The $Z_ k$ model
of  \cite{ZFa} amounts to choose the dimension of the parafermions
$\psi_n$ to be
$
h_{\psi_n} = {n(k-n)/k}$.
This is the simplest choice compatible
with the  monodromy  constraints.
A general solution to these constraints has also been displayed in  \cite{ZFa}. It reads
\begin{equation} \label{modelb}
h^{(\beta)}_{\psi_n} = {\beta n(k-n)\over k}+a_n\;,
\end{equation}
where $\beta$ is an integer and  $a_n$ are integers that satisfy $a_n=a_{k-n}$.  The second type of  generalization  is thus defined in terms of such a generalized form of the dimension of the parafermionic fields.

Actually, both types of generalization (with $a_n=0$ and $\beta$
positive) can be combined,  leading to a much larger set of
parafermionic theories, denoted as $ Z_ k^{(g,\beta)}$, classified in terms
of both  their underlying structural algebra\footnote{More
precisely, the algebra $g$ is the finite form of the affine
algebra involved in the coset $\gh/\uh(1)^r$ pertaining to the
$\beta=1$ version, the algebra that governs the form of the
parafermionic OPE within this class.}  $g$ and the parameter
$\beta$.   In this notation, the original parafermionic theories
of \cite{ZFa} would correspond to the $ Z_ k^{(su(2),1)}$ models.


In this work we explore a generalization of the second type. The
corresponding models will be referred to as the $Z_ k^{(su(2),\beta)}$
models, written  $Z_ k^{(\beta)}$ for short. Some $Z_ k^{(\beta)}$ models have been
considered previously. The unitary sequence of the $ Z_3^{(2)}$ models, first introduced
in appendix A of \cite{ZFa}, have been studied in more
detail in  \cite{ZFc}. Further developments are presented in \cite{FPT}.  More recently, the unitary sequence of the $ Z_k^{(2)}$ models for arbitrary $k$ have been analyzed in depth in \cite{Dot}. The only known results for higher values of $\beta$ seems to be those presented in \cite{DotS} pertaining to $\beta=4$ and $k=3$.

Here we explore a novel possibility, which is to consider $\beta$ to be half-integer. This is allowed whenever $k$ is even.  The possibility of having $\beta$ non-integer seems to have first been  mentioned in \cite{Rav}, although it has  not  been studied there. This enhancement in the range of the possible values of $\beta$ in relation with the parity of $k$   is a direct consequence of the requirement   $k\beta\in \Z$. This, in turn, ensures that moving the parafermionic field $\psi_1(z)$ $k$ times around $\psi_1(w)$ should not produce any phase.

The special case $\beta=3/2$ is analyzed here in some detail.\footnote{This turns out to be  the simplest possibility when $\beta$ is positive since the associativity conditions are not satisfied for $\beta=1/2$.} Since the dimension of the  basic parafermion reads
\begin{equation}
h^{(3/2)}_{\psi_1}= {3\over 2}\left(1-{1\over k}\right) \; , \end{equation}
this provides a sort of parafermionic deformation of supersymmetry.\footnote{Exotic supersymmetry has been explored in \cite{Rav}  also from a parafermionic point of view. However, the models considered there are such that the basic parafermionic field has dimension $1+1/k$, which requires $\beta=-1$ and $a_1=2$. This choice of dimension is motivated by the aim of reproducing the relation $Q^k \sim P$ between the exotic supercharge $Q$ and the translation operator $P$.  }

Like for the $Z_k^{(1)}$ models,  an associativity analysis
is enough to fix the central charge of the $Z_ k^{(3/2)}$
models  to
\begin{equation}\label{centre}
c=-{3(k-1)^2\over (k+3)}\; . \end{equation}
For $k=2$, the central charge is $-3/5$,  which is precisely that of the  Virasoro minimal model $ {\cal M}(3,5)$. The  $Z_
2^{(3/2)}$  model is indeed equivalent to ${\cal M}(3,5)$.\footnote{Note also that $ {\cal M}(3,5)$ is  the simplest graded parafermionic model, associated to the coset ${\osp}(1,2)_\kappa/\uh(1)$ for $\kappa=1$ \cite{JM}.} This relation between the $Z_
2^{(3/2)}$ model and a particular Virasoro minimal model is the simplest example of a general relationship between $Z_
k^{(3/2)}$ and  $W_k$ models, whose precise phrasing is
\begin{equation}\label{zwiden}
Z_
k^{(3/2)}\simeq W_k^{(k+1,k+3)} \;.\end{equation}
Note that for  $k$ even (and positive), $k+1$ and $k+3$ are ensured to be relatively prime.


\section{Structure of the $Z_
k^{(3/2)}$  algebra}

Consider the set of $k$ parafermionic fields $\psi_n$, $n=0,\cdots,  k-1$, of conformal dimensions
\begin{equation}
h_n\equiv h^{(3/2)}_{\psi_n}= {3\over 2} n \left(1-{n\over k}\right)\; . \end{equation}
As usual with parafermions, the mode decomposition of $\psi_1$ depends upon the sector on which
it acts. These sectors are labeled by
integers $t=0,\cdots , k-1$. In the sector $t$, we have
\begin{equation}
 \psi_1(z)  = \sum_{m=-\y}^\y
z^{-t/k-m-1}A_{m+1+t/k-3/2+3/2k}=\sum_{m=-\y}^\y
z^{-\la q-m-1}A_{m-1/2+\la(1+q)} \; , \label{modep}\end{equation}
i.e., the sector determines  the fractional power of $z$. In the
second relation, we
have traded
$t$ for its rescaled version
$q$, and wrote $3/2k$ as $\la$:
\begin{equation}
t={3 q\over 2}\; , \qquad \la= {3\over 2k}\; . \end{equation}
A similar
expression holds for the decomposition of
  $\psi^\dagger_1$, with $q\rw -q$. 
The
  inverted versions of (\ref{modep}) and its dagger version are:
  \begin{equation} \label{modepin}
   A_{m-1/2+\la(1+q)} = {1\over 2 \pi i}\oint_0 dz\,
    z^{\la q + m}\,\psi_1 (z)\,
  \qquad {\rm and} \qquad
   A_{m-1/2+\la(1-q)}^\dagger  ={1 \over 2 \pi i} \oint_0
{dz }\,  z^{-\la q+m}\,\psi^\dagger_1 (z)\;.
\end{equation}

Interpreting $q$ as the charge that
characterizes the sector $t$ amounts to  assign the charge $q=2$
to $A$ (actually  this is essentially the reference charge, the
charge concept being relative)
while that of $A^\dagger$ is $-2$. It is convenient to drop the
fractional part (more precisely, the part  proportional to $\la$)
since it is easily reconstructed from the charge of the state on
which the mode acts), signaling this omission in the mode writing
by replacing $A$ by $\A$. Thus, when acting on an arbitrary  state
of charge $q$, we write
\begin{equation}
A_{u+\la(1+q)}= \A_u\end{equation}

The defining (holomorphic) OPEs of the $Z_
k$ parafermionic conformal algebra \cite{ZFa}
are
\begin{equation}
\begin{split}
\psi_n (z) \,\psi_{n'} (w) &\sim {c_{r,s}\over  (z-w)^{3nn'/k}}\;
 \psi_{n+n'} (w)  \qquad (n+n'<k) \\
 \psi_n(z) \,\psi^\dagger_{n'} (w) &\sim { c_{n,k-n'}\over
(z-w)^{3{\rm min}(n,n')-3nn'/k} }\;
  \psi_{k+n-n'} (w)  \qquad (n+n'<k) \\
\psi_n(z) \,\psi^\dagger_n (w) &\sim {1\over (z-w)^{3n(k-n)/k}}
\left[I+(z-w)^2 {2 h_{n} \over c}\, T(w) +\cdots\right]\;,
\end{split}
\end{equation}
where  $\psi_0=I$ and $ \psi^\dagger_n=
\psi_{k-n}$.
The remaining OPEs are obtained by conjugation, with
$c_{n,n'}= c_{k-n,k-n'}$.

Following \cite{ZFa}, the  commutation relations are found to be
\begin{equation} \label{comisaa}
  \sum_{l=0}^{\infty} C^{(l)}_{-2\la} \left[ \A_{n-l-1/2}
\A^\dagger_{m+l+1/2} - \A_{m-l+1/2}^\dagger  \A_{n+l-1/2}
  \right]
  = \left[ {2h_{1}\over c}  L_{n+m} + {1 \over 2}
(n+{\la q})(n-1+\la q) \delta_{n+m,0} \right]
\end{equation}
 where (with $c$ given by (\ref{centre}))
\begin{equation}
  {2h_{1} \over c }= - {(k+3)\over k(k-1) } \;   \qquad {\rm and}\qquad  C^{(l)}_a={\Gamma(l-a)\over \Gamma(-a)
l!}\;. 
 \end{equation}

In the same way, we easily obtain the following commutation
relations involving only $\A$ modes:
\begin{equation}
  \sum_{l=0}^{\infty} C^{(l)}_{2\la} \left[ \A_{n-l-1/2}
\A_{m+l+1/2} - \A_{m-l+1/2}  \A_{n+l-1/2}
  \right]  = 0 \;.
\label{comisaTwo}\end{equation} There is an identical expression
with $\A$ replaced by $\A^\dagger$ \footnote{The relative sign in
(\ref{comisaa}) indicates that $\psi_1$ and $\psi_1^\dagger$ are
mutually fermionic. The one in (\ref{comisaTwo}) shows that $\psi_1$
is not fermionic with respect to itself. These signs have been
obtained by associativity.}.

The highest-weight states $|{\rm hws}\R$ are naturally defined from being annihilated by the action of the positive parafermionic modes:
\begin{equation}
\A_{m+1/2}|{\rm hws}\R  = \A_{m+1/2}^\dagger|{\rm hws}\R= 0 \quad
{\rm for}\,\;\; m\geq 0\;. \label{highestW}\end{equation} The
vacuum $|0\R$ is  certainly a particular example of a
highest-weight state. The parafermionic  field $\psi_n$  itself
is  associated to the following vacuum descendant state:
\begin{equation}
\psi_n(0)|0\R= (\A_{-3/2})^n|0\R\;.
\end{equation}


\section{Structure constants and central charge}

In this section, we will fix the value of the central charge and the structure constants for the $Z_
k^{(3/2)}$ parafermionic theory.  This will be done by a rather simple method using mode computations (cf. \cite{JM3p}).

Let us start by fixing the central charge.  The trick is to start with a simple string of three modes acting on the vacuum. For the first mode $\A_u|0\R$ we chose the highest value of $u$ that makes the state non-vanishing. This is $u=-3/2$. The second mode is chosen to be such that the resulting state  is proportional to $L_{-1}|0\R$. This second mode is thus $\A^\dagger_{-1-u}= \A^{\dagger}_{1/2}$. Finally, we choose the third mode such that the  resulting state  is proportional to $\A_u|0\R$. We thus consider
\begin{equation}
\A_{u+1}\A^\dagger_{-1-u} \A_{u}|0\R=\A_{-1/2}\A^{\dagger}_{1/2}\A_{-3/2} |0 \rangle \;.
\end{equation}
By commuting the two rightmost term (which is indicated below by the underbrace) with (\ref{comisaa}), we have
 \begin{equation}
 \A_{-1/2} \underbrace{\A^{\dagger}_{1/2}\A_{-3/2}}|0 \rangle =0\;. \end{equation}
This, of course, is compatible with the fact that $\A^{\dagger}_{1/2}\A_{-3/2}|0 \rangle$ is proportional to
 $L_{-1}|0\R=0$.
On the other hand, by commuting the two left-most terms still using (\ref{comisaa}), we obtain
\begin{equation}
\begin{split}
\underbrace{\A_{-1/2}\A^{\dagger}_{1/2} } \A_{-3/2} |0 \rangle &=\, {2 h^2_{1} \over c}\A_{-3/2} |0 \rangle + {1 \over 2} (2 \lambda)(-1+2 \lambda)\A_{-3/2} |0 \rangle - C^{(1)}_{-2\lambda} \A_{-3/2}\A^{\dagger}_{3/2}\A_{-3/2} |0 \rangle \cr
&= \left( {9 \over 2}{ (k-1)^2 \over c k^2} + {3 \over 2} {(3-k) \over k^2} +{3 \over k}\right)\A_{-3/2} |0 \rangle \;.
\end{split}
\end{equation}
In establishing this result, we used  the case $n=1$ of the following relations:
\begin{equation}
\A^{\dagger}_{3/2+\ell}(\A_{-3/2})^n |0 \rangle= 0= \A_{-3/2+\ell}(\A_{-3/2})^n|0 \rangle\qquad {\rm for }\qquad \ell>0\;, \label{restri}\end{equation}
 which are easily proved by means of the commutation relations (\ref{comisaa}) and (\ref{comisaTwo}) supplemented by an inductive argument. Coming back the computation of the central charge, equating the two expressions obtained for $\A_{-1/2}\A^{\dagger}_{1/2}\A_{-3/2} |0 \rangle$  yields the announced result (\ref{centre}).
This value can be recalculated in several different  ways, confirming thereby the associativity of the theory.

Let us now turn to the computation of the structure constants. We first show that the constants $c_{n,n'}$ are fixed by $c_{1,n}$. Introducing the compact  notation
\begin{equation}
(c)_n= c_{1,1}\cdots c_{1,n-1}\;, \end{equation}
we have, in a schematic notation  (i.e., by dropping the $z$ dependence and regarding these equalities in terms of leading terms):\begin{equation}
\psi_{n+n'}= {1\over (c)_{n+n'}}(\psi_1)^{n+n'}= {(c)_n(c)_{n'} \over (c)_{n+n'}}\psi_{n}\psi_{n'}=
{(c)_n(c)_{n'} \over (c)_{n+n'}} c_{n,n'}\psi_{n+n'}\;.\end{equation}
{}From this, we conclude that
\begin{equation}
c_{n,n'}= { (c)_{n+n'}\over (c)_n(c)_{n'} }\;, \label{reCu}\end{equation}
the sought for  relationship. Note also that
\begin{equation}
\psi_1^\dagger (\psi_1)^{n+1}= (c)_{n+1} \psi_1^\dagger \psi_{n+1} =  (c)_{n+1} c_{1,n}\psi_{n}= c_{1,n}^2(\psi_{1})^n\;. \end{equation}
We are now in position to calculate $c^2_{1,n}$. From the previous relation, we get
\begin{equation}
\A^{\dagger}_{3/2}(\A_{-3/2})^{n+1} |0 \rangle = c^2_{1,n} (\A_{-3/2})^{n} |0 \rangle\; . \end{equation}
We next commute $\A^{\dagger}_{3/2}$ with the first $\A_{-3/2}$ factor using (\ref{comisaa}):
\begin{equation}
\begin{split}
\A^{\dagger}_{3/2}(\A_{-3/2})^{n+1} |0 \rangle &= \left[-{2h_1\over c}L_0 - {1 \over 2}{ (-1+2n\la) (-2+ 2n \lambda) } \right](\A_{-3/2})^{n} |0 \rangle +\A_{-3/2}\A^{\dagger}_{3/2}(\A_{-3/2})^{n} |0 \rangle
\cr
&\equiv \,  \Delta_n \, (\A_{-3/2})^{n} |0 \rangle +\A_{-3/2}\A^{\dagger}_{3/2}(\A_{-3/2})^{n} |0 \rangle
\;,
\end{split}
\end{equation}
where $\Delta_n$ stands for
\begin{equation}
\Delta_n =
-{2h_1h_n \over c} - {1 \over 2}{ (-1+2n\la) (-2+ 2n \lambda) } \;, \end{equation} and used
\begin{equation}
L_0(\A_{-3/2})^{n} |0 \rangle = h_n (\A_{-3/2})^{n} |0 \rangle\;.  \end{equation}
 In the intermediate steps, the conditions  (\ref{restri}) have been taken into account.
Now by iterating this result, we get, with  $c$ fixed as above,
\begin{equation}
\A^{\dagger}_{3/2}(\A_{-3/2})^{n+1} |0 \rangle= \left[ \sum_{j=0}^n \Delta_j \right](\A_{-3/2})^{n} |0 \rangle=
- {(n+1)(k-n)(k-2n-1) \over k (k-1)}(\A_{-3/2})^{n} |0 \rangle \;.  \end{equation}
Comparing the two distinct expressions we have derived for this state, we end up with
\begin{equation}
c^2_{1,n} = - {(n+1)(k-n)(k-2n-1) \over k (k-1)}\;.
\end{equation}
Note that the generic structure constant have to be imaginary. In the simplest case $k=2$, we see that $c_{1,1}^2=c_{1,k-1}^2=1$ as it should.
Finally, using (\ref{reCu}), we have

\begin{equation}
c^2_{n,n'} = - {(n+n')!\, (k-n)! \, (k-n')!\, (k-2n-1)!! \,
(k-2n'-1)!! \over k(k-1)\, (k-1)!\, n! \, n'! \, (k-n-n')!\,
(k-3)! !\, (k-2n-2n'-1)!!}\;.
\end{equation}


\section{Primary singular vectors and spectrum}

Verma modules are obtained by acting on the highest-weight states with the different  lowering operators. A spanning set of states is given by:
%
\begin{equation}
\begin{split}
& L_{-n_1} \cdots L_{-n_{m_1}} \A_{-1/2-n'_1} \cdots
\A_{-1/2-n'_{m_2}} \A^{\dagger}_{-1/2-n''_1} \cdots
\A^{\dagger}_{-1/2-n''_{m_3}} |{\rm hws}\R \cr & {\rm for}  \cr
 & n_i \geq n_{i+1} \geq 1 \; ;
\;\; n'_i \geq n'_{i+1} \geq 0 \; ; \;\; n''_i \geq n''_{i+1} \geq
0 \;,\cr
\end{split} \label{verma}
\end{equation}
with $m_1,\, m_2$ and $m_3$ running over all integers. Note that the above states have relative charge $2(m_2-m_3)$. We are interested in finding the characteristics of those highest-weight states that make the modules completely reducible, that is, for which there is an infinite number of the above states that are not independent.

Like for any parafermionic theory, if a string made of $\A_{-1/2}$
or $\A^{\dagger}_{-1/2}$ modes, acting on a highest-weight state, is
allowed to run freely, it will eventually reach a conformal
dimension lower than that of the highest-weight state.  (We stress
that in order to see this, we need to take into account the
fractional part of the modes.) For that reason, it is natural to
look for singular vectors in the form of $(\A_{-1/2})^{r+1}|{\rm
hws}\R$ and $(\A^{\dagger}_{-1/2})^{r'+1}|{\rm hws}\R$ for some integers $r$ and $r'$.  Yet, we
have no characterization of the highest-weight states. For sure,
it must belong to a definite sector $t$ (and recall for the usual
parafermions, the sector label characterizes the highest-weight
state uniquely).

So we first require the singular vectors
\begin{equation}\label{singA}
(\A_{-1/2})^{r+1}|{\rm hws}\R = 0
\end{equation}
to obey the highest-weight conditions (\ref{highestW}).  This
fixes the conformal dimensions  of the highest-weight states to be
\begin{equation}
h_{t,r} = -{ k (k-2r-t-1)(2r+t) +t^2 \over 2k(k+3)}\;.
\label{confDim}\end{equation} We next look for singular
vectors in the form $(\A^{\dagger}_{-1/2})^{r'+1}|{\rm hws}
\rangle $ leading the to same conformal dimension for the highest-weight states.
We thereby obtain
\begin{equation}
( \A^{\dagger}_{-1/2} )^{r+t+1} |{\rm hws} \rangle =0 \; .
\label{singB}\end{equation}

It is clear at this point  that we cannot get rid of this second
parameter $r$.  Consequently, it must become a second quantum
number characterizing our highest-weight states:
\begin{equation}|{\rm hws}
\rangle \equiv |t,r\rangle \;. \end{equation}

At this stage, the absence of any dependence upon $k$,  meaning
that the different  models would have the same primary singular
vectors,  indicate us that we certainly do not have obtained the
full set of primary singular vectors that would characterize the
completely reducible representations.  In order to unravel further
constraints, observe that we have yet no way of removing states of
the form $(A_{-3/2})^\ell (A_{-1/2})^{r} |t,r \rangle$, which, for
$\ell$ sufficiently large,  will once again have negative relative
conformal dimension. It is thus unsurprising  to discover another
set of null states that does depend upon $k$:
\begin{equation}
( \A_{-3/2} )^{k-t-2r+1} ( \A_{-1/2} )^{r} |t,r \rangle =0\;,
\label{singD}\end{equation}
and
\begin{equation}
( \A^{\dagger}_{-3/2} )^{k-t-2r+1}( \A^{\dagger}_{-1/2} )^{r+t} |t,r \rangle =0\;.
\label{singF}\end{equation}
Note however that these null states  $|\chi \rangle$ have been obtained as the solutions of the conditions
\begin{equation}
\A_{3/2} |\chi \rangle= 0 = \A^{\dagger}_{-3/2} |\chi \rangle
\end{equation}
instead of (\ref{highestW}).  These conditions are enough to
ensure their decoupling from the whole module, but they also
suggest that these states do arise as descendants of  genuine
singular vectors.  We found the simplest of such singular vectors
in the module of relative charge $0$ to be:
\begin{equation} \label{mixe}
\left[L_{-1} + { 2(2r+t)(2r+t-1) \over (t-1)(2r+t+3) } \A_{-1/2} \A^\dagger_{-1/2}  \right]|t,r
\rangle= 0 \quad{\rm for }\quad k=2r+t\;.
\end{equation}
This is indeed a  primary singular vector,  obeying the highest-weight
conditions (\ref{highestW}).
 By acting on this vector (\ref{mixe}) with $(A_{-1/2} )^{r+1}$ and taking (\ref{singA}) into account, we get
 \begin{equation}
 \A_{-3/2}  ( \A_{-1/2} )^{r} |t,r \rangle =0\;,
\label{singDD}\end{equation} which is identical to  (\ref{singD})
for $k=2r+t$.  We can also generate (\ref{singD}) in a similrar way. We thus naturally expect a whole sequence of
similar singular vectors to be at the source of (\ref{singD}) and
(\ref{singF}).\footnote{The complete singular-vector structure and
the resulting character formula will be presented elsewhere.}


By definition, $r$ has to be a non-negative integer.  From (\ref{singD}) or  (\ref{singF}), we deduce that $0 \leq r \leq
(k-t)/2$.  This bound, together with $0\leq t\leq k-1$, 
allow us to fix the spectrum of the theory.




\section{Conclusion}

We have introduced new $Z_
k$ parafermionic models that belongs to the general class of models introduced in \cite{ZFa}: the structure of
the algebra is still a cyclic  $Z_k$-one but the dimension of the basic parafermion is  modified as compared with
those of the usual $\suh(2)_k/\uh(1)$ model, by the multiplicative factor $3/2$.   In this letter, we have confined
ourself to the study of  the  basic properties of these models, such as the determination of the essential parameters
of the theory (structure constants and central charge) as well as a determination of the spectrum of the reducible modules.
Much remains to be done and/or clarified: the  determination of the field identifications, the complete analysis of the
singular vectors, the derivation of character formulae, the unravelling of  a quasi-particle basis and the corresponding
fermionic form of the character, etc. These topics will be considered elsewhere. Our main aim here was to establish the
well-definiteness of these models. Further support for this  comes from their proposed identification with the $W_k(k+1,k+3)$ models, which has been explicitly checked for $k=2,4,6,8$.\footnote{Note that the central charge of the $W_k^{(p',p)}$ models is
$$
c= (k-1)\left(1-{k(k+1)(p-p')^2\over pp'}\right)
$$
and with $(p',p)= (k+1,k+3)$, this reduces to (\ref{centre}).}

With regard to the last point, let us indicate that the identification (\ref{zwiden})  can be generalized in a
very natural way, for all integer or half-integer values of $\beta\geq 1$ as follows:
\begin{equation}\label{gzwiden}
Z_k^{(\beta)}\simeq W_k^{(k+1,k+2\beta)}\;. \end{equation}
 For $\beta=3/2$, this reduces to
(\ref{zwiden}) while for $\beta=1$, this is the standard
identification of the $Z_k^{(1)}$ models as the simplest
minimal $W_k$ model \cite{Nar}.
 As evidence for (\ref{gzwiden}) when $\beta>
3/2$, we notice that the dimension  $W_k$ field labeled by the two $\suh(k)$ weights
$\{{\widehat\omega}_0,\muh\}$ (at respective levels 1 and 3) and such that the corresponding non-zero finite weight is
$\mu= 2\beta \om_1$ matches precisely that of  $\psi_1$, namely
$\beta (1-1/k)$. We also verified that under some restrictions,
associativity  fixes the central charge of the
$Z_3^{(2)}$ model to that of the $W_3^{(4,7)}$ model.

Let us stress that for $\beta=2$,  the relation between the $Z_k^{(2)}$ models and the $W_k^{(k+1,k+4)}$ models is not recovered in  \cite{Dot}.  The reasons for this might be that the
authors have focussed on unitary solutions
and/or finding solutions for which the conformal dimension of
the primary fields $\phi_t$ is  symmetric under
the transformation $t\rightarrow k-t$ (as for the usual
parafermions). We note that the primary fields do not have this
kind symmetry for the $Z_k^{(3/2)}$ models (cf. (\ref{confDim})). This relation however, is not incompatible with the associativity analysis. For instance, the expression for  the $Z_k^{(2)}$  central charge, parameterized in terms of a number $\la$,  reads (cf. eq. (A.8) of \cite{ZFa})):
\begin{equation}\ c_{Z_k^{(2)}}= {4(k-1)(k+\la-1)\la\over (k+2\la)(k+2\la-2)}\;,
\end{equation}
while that of  the   $W_k^{(k+1,k+4)}$ models is:
\begin{equation}
c_{W_k^{(k+1,k+4)}}=(k-1)\left(1-{9k\over (k+4)}\right)= -{4(k-1)(2k-1)\over (k+4)}\;.
\end{equation}
Enforcing the equality of these two expressions yields the two solutions\footnote{For $k=2$,  for instance, $\la=0,-1$. With  $\la=0$ say, we set $k=2+\e$ so that $\la=-\e/3$ et thus:
$$
c^{\rm ZF}= {4(1+\e)(1+2\e/3)(-\e/3)\over (2+\e/3)(\e/3)}\xrightarrow{(\e\rw0)}-2
$$
which corresponds to that of  $\M(3,6)=\M(1,2)$. On the other
hand, without this connection between $\la$ and $k$, we get $c=1$
by directly setting  $k=2$ in the expression for $c_{Z_k^{(2)}}$,
thereby confirming that both solutions are possible in this case.}
\begin{equation}
\la= {(2-k)\over 3} \quad {\rm or} \quad {(1-2k)\over 3} \;. \end{equation}
The relationship with the work \cite{Dot}   certainly requires further analysis but (\ref{gzwiden})
suggests that the sequence $W_k^{(k+1,k+4)}$ corresponds to the first of an infinite sequence of non-unitary solutions for an
eventual complete realization of the $Z_k^{(2)}$ models.

Note finally that  for $k=2$, (\ref{gzwiden})  reduces to
$
Z_2^{(p/2-1)}\simeq \M(3,p).$
The dimension of $\psi_1$, namely $(p-2)/4$, is precisely that of $\phi_{2,1}$. The corresponding $Z_
2$ parafermionic description of these Virasoro minimal models has been studied recently in \cite{JM3p}.
The relation (\ref{gzwiden}) hints for the existence of a similar parafermionic description of these non-unitary $W_k(k+1,k+2\beta)$ models.



\noindent {\bf ACKNOWLEDGMENTS}

The work of PJ is supported by EPSRC and partially  by the EC
network EUCLID (contract number HPRN-CT-2002-00325), while that of  PM is supported  by NSERC.

\end{document}